\documentclass{article}

\usepackage{apacite}
\usepackage[english]{babel}

\usepackage[letterpaper,top=2cm,bottom=2cm,left=3cm,right=3cm,marginparwidth=1.75cm]{geometry}

\usepackage{amsthm}
\usepackage{verbatim,color,amssymb}
\usepackage{amsmath}					
\usepackage{amsthm}					
\usepackage{algorithm,algorithmic}
\usepackage{natbib}
\usepackage{setspace}
\usepackage[mathscr]{euscript}
\usepackage{fancyhdr}
\usepackage{enumitem}
\usepackage{graphicx}
\usepackage{geometry}
\usepackage{lineno}
\usepackage[compact]{titlesec}
\usepackage{listings}
\usepackage{rotating}
\usepackage{booktabs}
\usepackage{subcaption}
\usepackage{mathtools}
\usepackage[utf8]{inputenc}
\usepackage[T1]{fontenc}
\usepackage{float}
\usepackage{tikz}
\usetikzlibrary{arrows,chains,backgrounds,fit}
\usepackage{multirow}
\usepackage{lineno}

\newtheorem*{Proof*}{Proof}

\def\log{\hbox{log}}

\def\MVN{\hbox{MVN}}

\def\log{\hbox{log}}

\def\Invwish{\hbox{Inv-Wishart}}

\def\P_25_ICML{{\it Proceedings of the 25th international conference on Machine learning}}

\def\bse{\begin{eqnarray*}}
	\def\ese{\end{eqnarray*}}
\def\be{\begin{eqnarray}}
\def\ee{\end{eqnarray}}
\def\bq{\begin{equation}}
\def\eq{\end{equation}}

\def\b1e{{\mathbf e}}

\def\bq{{\mathbf q}}

\newcommand{\uiota}             {\mbox{\boldmath$\uiota$}}

\newcommand{\mbf}[1]{\mbox{\boldmath $#1$}}

\graphicspath{ {/Plots/} }
\usepackage{amsmath}
\usepackage{graphicx}
\usepackage[colorlinks=true, allcolors=blue]{hyperref}

\title{A Bayesian Functional Accelerated Failure-Time Model with Varying Effects Correcting for Measurement Error}
\author{Joseph Yang, Roger S Zoh, Carmen D Tekwe, Lan Xue}

\begin{document}
\maketitle

\begin{abstract}

Functional data collected as continuously observed trajectories arise naturally in many biomedical settings, and a key inferential goal is understanding how such functional covariates relate to time-to-event outcomes while allowing that relationship to vary across subgroups defined by scalar characteristics. Existing frequentist approaches to functional accelerated failure-time (AFT) models struggle to flexibly capture the joint, nonlinear influence of scalar covariates and time on the functional effect, and none adequately address the measurement error that frequently contaminates functionally observed exposures. We propose a Bayesian functional AFT model in which the functional coefficient is a varying effect function of both time and a set of scalar covariates, modeled through a Gaussian process single-index structure that provides a flexible, nonlinear framework for subgroup modification. Measurement error in the functional covariate is handled by pairing a proxy observation with an instrumental variable that accommodates non-linear associations with the latent functional exposure. A prominent source of such functional data is wearable devices, which can continuously monitor physical activity (PA) behavioral patterns — such as step counts — over time, yet whose outputs are well known to be prone to measurement error and to exhibit heterogeneous associations with health outcomes across demographic subgroups. Through simulations, we show that our approach recovers the true varying functional effects and reduces bias relative to naïve models that ignore measurement error. We apply our methods to the Reasons for Geographical and Racial Differences in Stroke (REGARDS) study to investigate how step-count physical activity relates to time-to-death from ischemic stroke across racial and regional groups.

\end{abstract}


\section{Introduction} \label{s:intro}

In many scientific domains where predictors are observed as trajectories over a continuous scale, functional data analysis has become a widely used framework with applications in health monitoring, environmental exposures, and longitudinal behavioral studies. A central objective in these settings is to understand how functional covariates relate to scalar outcomes, including time-to-event responses in epidemiological and biomedical research. Functional data survival analysis, including scalar-on-function regression, enables investigation of the association between function-valued trajectories and time-to-event scalar outcomes, with early developments based on Cox proportional hazards (PH) models with functional time-varying predictors (\citealt{muller2005generalized}). Subsequent work has extended these ideas using penalized likelihoods (\citealt{goldsmith2011penalized}) and functional generalized additive models (\citealt{mclean2014functional}). While the functional Cox PH model is commonly used as a flexible semi-parametric method, functional accelerated failure-time (AFT) models offer a parametric alternative that directly models survival time and can yield improved fits when the proportional hazards assumption is violated (\citealt{gencer2025lung}). Efficient estimation techniques and asymptotic inference for functional AFT models have also been developed by \cite{liu2025efficient}, while related work has demonstrated the practical utility of functional AFT models in the application of longitudinal biomarkers (\citealt{qian2025functional}). 

In practice, functional covariates are often subject to uncertainty arising from imperfect observation processes, including device mis-calibration and incomplete recordings. As a result, the observed functional covariate may be viewed as a proxy for an unobserved latent covariate. One principled approach to addressing this uncertainty is through measurement error modeling, where studies have suggested that failure to account for such discrepancies can induce bias in model parameters (\citealt{Carroll2006}). Methods for correcting measurement error in covariates include regression calibration (\citealt{spiegelman1997regression}, \citealt{wang2021semiparametric}), but can be computationally costly when extended to high-dimensional functional data. Mixed-effects model-based (MEM) approaches address this through two-stage procedures in which the latent functional covariate is first estimated and then incorporated into the outcome model (\citealt{luan2023scalable}). 
Furthermore, Bayesian hierarchical models offer a flexible alternative by explicitly modeling the functional covariate as a latent process, allowing both the relationship between the outcome and the latent process, as well as the relationship between the observed surrogate and the latent covariate, to be modeled flexibly within a unified framework (\citealt{sinha2010semiparametric}).  

In addition to measurement error, the effect of functional covariates may vary across individuals depending on observed characteristics. In many biomedical and epidemiology studies, 
the association between physical activity trajectories and health outcomes may vary across individuals due to underlying heterogeneity in population characteristics. 
Assuming a common population-level functional effect in such settings can obscure important effect modification and lead to oversmoothed or misleading inference. These considerations have motivated the development of varying-coefficient models that allow covariate effects to vary smoothly with modifying variables, beginning with the work by \cite{hastie1993varying}. More recently, Bayesian approaches have emerged as flexible alternatives for modeling heterogeneous effects. For example, Bayesian tree-based varying-coefficient models proposed by \cite{deshpande2020vcbart} represent regression effects through additive ensembles of trees. In the context of functional data analysis, \cite{guhaniyogi2022distributed} developed Bayesian approaches to varying coefficient models with Gaussian process priors that allow regression effects to smoothly vary with scalar covariates. \cite{wu2010varying} provide a foundational frequentist approach to varying-coefficient functional regression, where the regression surface is modeled as a smooth function of the modifying covariate and employs functional principal component analysis with local polynomial smoothing to estimate covariate-specific effects. While this method demonstrates improved interpretability and predictive performance, it is limited to least-squares regression for continuous outcomes and does not accommodate censoring in time-to-event responses and measurement error in functional covariates within a unified probabilistic model. 

In this work, we propose a Bayesian scalar-on-function accelerated failure-time model that jointly addresses censoring, measurement error in functional covariates, and covariate-dependent heterogeneity in functional effects. To date, most of the existing methodologies above address these challenges in isolation. Our modeling approach integrates a flexible single-index varying-effect structure with measurement error correction using an observed proxy and an instrumental variable. This unified framework enables coherent posterior inference for subgroup-specific functional effects while reducing bias induced by measurement error. 

\section{Model Approach}
\subsection{Model Specification}

We consider right-censored survival data where $S_i$ and $C_i$ denote the true survival and censoring time respectively for individuals $i=1, \dots, n$. Suppose each individual is observed with survival time $T_i = \min(S_i, C_i)$, error-free scalar covariates $\{\mbf{Z}_{1i}, \mbf{Z}_{2i}\}$, and $W_i(t), t\in[0, 1]$, then the scalar-on-function accelerated failure-time model with functional covariates subject to measurement error is 
\begin{align}
\log(S_i) &= \beta_0 + \mbf{Z}_{1i}^\top\mbf{\beta}_1 + \int_0^1\beta_2(\mbf{Z}_{2i}, t)X_i(t)dt + \sigma_{\varepsilon}\varepsilon_i \label{eq:1}\\ 
W_i(t) &= X_i(t) + U_i(t).  \label{eq:2}
\end{align}
The coefficient $\beta_2(\mbf{Z}_{2i}, t)$ is a function of both time $t$ and the subject-specific scalar covariates $\mbf{Z}_{2i}$, thereby accommodating heterogeneity in the functional effect of $X_i(t)$ across individuals. 
We assume $\varepsilon_i$ are i.i.d. and follow distributions commonly used in accelerated failure time models (e.g. normal, logistic, exponential, etc). $W_i(t)$ is an unbiased proxy of $X_i(t)$ observed with measurement error that is modeled as a mean-zero Gaussian process $U_i(t)$ with $cov(U_i(t), U_i(t')) = \sigma_u(t, t') < \infty$. Following   previous work on measurement error modeling (\citealt{sarkar2018bayesian} and \citealt{zoh2024bayesian}), the model is not identifiable based solely on the observed proxy $W_i(t)$, and thus requires additional assumptions to estimate the model parameters. Therefore, in addition to models \ref{eq:1} and \ref{eq:2}, we assume the following model for the observed instrumental variable (IV)) as 
\begin{align}
M_i(t) &= \delta(t)X_i(t) + \omega_i(t). \label{eq:3}
\end{align}
Similar to model \ref{eq:2}, we also assume that $\omega_i(t)$ is a mean-zero Gaussian process with $cov(\omega_i(t), \omega_i(t')) = \sigma_\omega(t, t')<\infty$. The scaling function $\delta(t)$ allows a nonlinear association between $M_i(t)$ and $X_i(t)$ over time due to systematic differences in measurement conditions, such as weekday and weekend activity patterns. 
For estimation, we scale the instrumental variable by $M_i^*(t) = \frac{M_i(t)}{\delta(t)}$ given that $\delta(t) \ne 0$ so that the model in \eqref{eq:3} yields
\begin{align}
M_i^*(t) = X_i(t) + \omega_i^*(t) \label{eq:4}
\end{align}
where $\omega_i^*(t) = \omega_i(t)/\delta(t)$. 

\subsection{Dimension reduction}

To reduce computational complexity and enable flexible modeling of effect heterogeneity, we adopt the basis expansion and single-index structure proposed by \cite{li2010generalized}. More specifically, we preselect $K$ 
basis functions $B(t) = \{b_1(t), \dots, b_K(t)\}^\top$ (e.g., B-splines, Fourier, or functional principal components) and represent the varying effect function as 
\begin{align}
\beta_2(\mbf{Z}_i, t) &= \sum^K_{k=1}b_k(t)\gamma_k(\mbf{\theta}^\top\mbf{Z}_{2i}) = \mbf{B}^\top(t)\mbf{\gamma}(\mbf{\theta}^\top\mbf{Z}_i). \label{eq:5}
\end{align}
$\mbf{\gamma}(\cdot) = (\gamma_1(\cdot), \dots, \gamma_K(\cdot))^\top$ are smooth univariate functions that vary with the single-index projection $\mbf{\theta}^\top\mbf{Z}_{2i}$, where the vector of weights $\mbf{\theta}$ lies on the unit sphere (i.e., $\lvert\lvert \mbf{\theta}\rvert\rvert = 1$) and is shared across all $K$ components, ensuring identifiability and reducing the multivariate dependence of $\mbf{Z}_{2i}$ to a single direction. 
Applying the same basis to the latent functional covariate $X_i(t)$, define the corresponding scores $X_{ik} = \int_SX_i(t)b_k(t)dt$, $U_{ik} = \int_SU_i(t)b_k(t)dt$, and $\omega_{ik} = \int_S\omega^*_i(t)b_k(t)dt$ and let $\mbf{X}_i = (X_{i1}, \dots, X_{iK})^\top, \mbf{U}_i = (U_{i1}, \dots, U_{iK})^\top$, and $\mbf{\omega}^*_i = (\omega^*_{i1}, \dots, \omega^*_{iK})^\top$. Substituting these expansions into models \eqref{eq:1}-\eqref{eq:4} gives the reduced-rank system
\begin{align}
\log(S_i) &= \beta_0 + \mbf{\beta}_1^\top\mbf{Z}_{1i} + \sum^K_{k=1}\gamma_k(\mbf{\theta}^\top\mbf{Z}_{2i})X_{ik} + \sigma\epsilon_i,  \label{eq:6}\\
\mbf{W}_i &= \mbf{X}_i + \mbf{U}_i,  \label{eq:7}\\
\text{and }\mbf{M}^*_i &= \mbf{X}_i + \mbf{\omega}^*_i \label{eq:8},  
\end{align}
and assume that $\mbf{X}_i \sim \MVN(\mbf{\mu}_x, \Sigma_x)$, $\mbf{U}_i \sim \MVN(\mbf{\mu}_u, \Sigma_u)$, and $\omega_i^*\sim \MVN(\mbf{\mu}_\omega, \Sigma_\omega)$. 

\subsection{Imputation of Censored Individuals}
In many works involving a Bayesian approach to fitting survival models, data augmentation is routinely leveraged to resolve complex censoring and thereby efficiently derive likelihood computations. See, for example, \cite{tanner1987calculation}. More specifically, by augmenting the model with latent log-event times for censored objects, we effectively bypass the survival-function term in the likelihood with a normal density evaluation truncated on the censoring threshold. In the conventional observed-data likelihood, censored observations require integration over the upper tail of the error distribution, which renders the model non-conjugate and greatly complicates MCMC updates. Following the data-augmentation approach in \cite{maity2020integration}, censored log-survival times are imputed via truncated normal distributions at each step within the MCMC posterior sampling (see supplementary material). 

\subsection{Model Assumptions and Priors}

We adopt a hierarchical Bayesian framework with priors specified for the model components. Diffuse priors are assigned to the fixed effects $\mbf{\beta}_1$, and the single-index weights $\mbf{\theta}$ are constrained to a unit sphere using a Fisher-von Mises prior (\cite{antoniadis2004bayesian}) for identifiability. The varying-effect functions $\gamma_k(\cdot)$ are modeled using a Gaussian process prior (\citealt{choi2011gaussian}) to capture smooth nonlinear dependence on the linear combination of scalar covariates $\mbf{\theta}^\top\mbf{Z}_{2i}$. Hyperparameters governing the covariance structure and smoothness of this Gaussian process are assigned with weakly informative priors similar to \cite{shi2005hierarchical} and \cite{neal2011mcmc}. For the measurement error model in equation \eqref{eq:7}, $\mbf{U}_i \sim \MVN(\mbf{\mu}_u, \Sigma_u)$ with $\mbf{\mu}_u \sim \MVN(\mbf{\mu}_{u, 0}, \Sigma_{u,0})$ and $\Sigma_u \sim \Invwish(\cdot \vert \nu_{u, 0}, \Psi_{u, 0})$. For the instrumental variable model in equation \eqref{eq:8}, $\mbf{\mu}_\omega \sim \MVN(\mbf{\mu}_{\omega, 0}, \Sigma_{\omega, 0})$ and $\Sigma_\omega \sim \Invwish(\cdot \vert \nu_{\omega, 0}, \Psi_{\omega, 0})$. 

\subsection{Posterior}

Posterior inference is conducted using a Metropolis-within-Gibbs sampler. The form of the conditional posterior distributions are outlined in 
section S.4 of
the supplementary material. To update the high-dimensional matrix of varying-effect coefficients, we employ the fast Gaussian sampling algorithm of \cite{bhattacharya2016fast}, which enables efficient sampling without explicit matrix inversion. The proposed Bayesian framework provides inference for the model parameters from the MCMC draws, and facilitates inference on subject-specific functional effects through the latent functional covariates $X_i(t)$. 

\section{Simulation}

To assess the performance of the proposed Bayesian functional varying-effects AFT model, we conducted a comprehensive simulation study using physical activity trajectories from the National Health and Nutrition Examination Survey (NHANES) cohort. Consistent with previous empirical simulation studies of functional accelerometry data (\citealt{parker2023bayesian}, \citealt{ghosal2023functional}), we generate datasets containing informative samples with realistic characteristics of the covariates, functional predictors, and model parameters are drawn directly from the NHANES cohort. Specifically, weekday and weekend average minute-level monitor-independent movement summary (MIMS) measurements were calculated for each individual. The latent functional covariate $X_i(t)$ was defined as the log-transformed $\log(\text{MIMS} + 1)$ weekday-average MIMS trajectory. From each individual, we also retained two continuous scalar covariates $\mbf{Z}_{1i}$ and a binary subgroup indicator $Z_{2i}$. 
True values of the functional effects corresponding to the two subgroups defined by $Z_{2i}$ were obtained by fitting general additive models with the \texttt{mgcv} package, and 
illustrated as dashed lines in Figure \ref{fig:Bestplot}.

\begin{figure}[h!]
    \centering
    \includegraphics[width=1\linewidth]{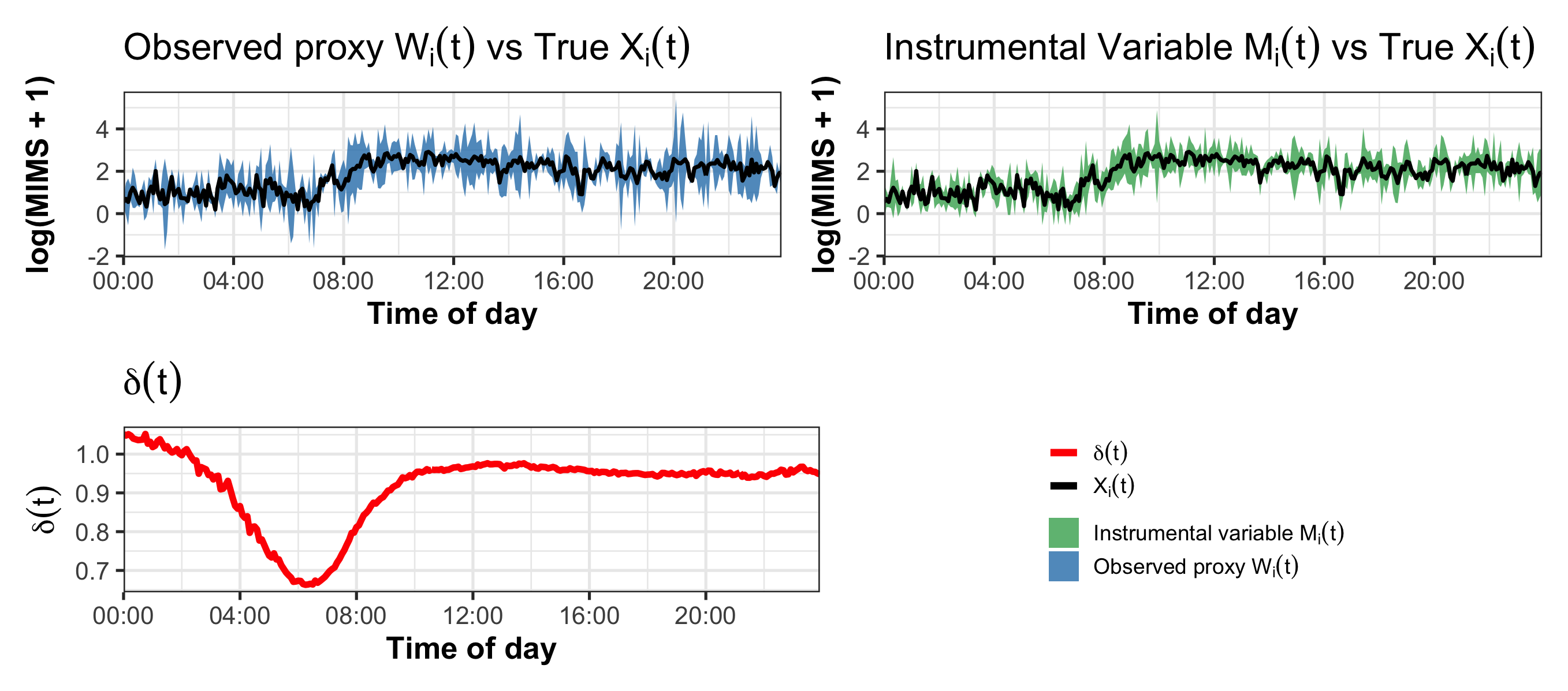}
    \caption{Simulated MIMS trajectories for one individual using NHANES Data}
    \label{fig:NHexample}
\end{figure}

We considered a factorial simulation design examining different sample sizes, measurement error noise, censoring levels, and correlation structures in the functional measurement error process. Specifically, we examined sample sizes $n \in \{300, 1000\}$ corresponding to small and large cohort settings, and simulated survival times
from the accelerated failure-time model $\log(S_i) = \beta_0 + \mbf{Z}_{1i}^\top\mbf{\beta}_1 + \int^{1}_{0} \beta_2(Z_{2i}, t)X_i(t)dt +\sigma_\varepsilon\varepsilon_i$, where $\varepsilon_i$ follows a normal distribution
and $\sigma_\varepsilon \in \{0.1, 0.5\}$. 
Right-censoring times were generated independently from an exponential distribution with the censoring rate calibrated with a root-finding algorithm similar to \cite{wan2017simulating} to achieve targeted censoring levels of approximately 40\% or 80\%. We simulated the observed proxy functional covariate as $W_i(t) = X_i(t) + U_i(t)$, where $U_i(t)$ is a mean-zero Gausssian process with variance $\sigma_u^2 \in \{0.5^2, 1.0^2, 4.0^2\}$ and 
correlation $\rho_u \in \{0, 0.25, 0.75\}$. 
The instrumental variable $M_i(t) = \delta(t)X_i(t) + \omega_i(t)$ is simulated based on weekend-average MIMS trajectories log-transformed similar to $X_i(t)$, and scaled by a time-varying function $\delta(t)$ defined as the ratio of weekend-to-weekday average MIMS. Similar to $U_i(t)$, $\omega_i(t)$ is a mean-zero Gaussian process with $\sigma_\omega = 0.5$ and correlation $\rho_\omega = 0.25$. An example of the functional covariates and $\delta(t)$ for a dataset simulated with $n = 300$ and $\sigma_u = 0.5$ is shown in Figure \ref{fig:NHexample}. 
Additionally, we considered simulation cases with increased confounding between the scalar covariates $\mbf{Z}_{1i}$ and latent functional covariate. In these cases, we induce dependency of $\mbf{Z}_{1i}$ through the basis coefficients used to construct $X_i(t)$, resulting in larger correlations between $\mbf{Z}_{1i}$ and $X_i(t)$ across time. 

We generated 800 replicated datasets and compare four modeling approaches: (i) an oracle MCMC approach assuming the latent function covariate $X_i(t)$ is fully observed without $W_i(t)$ and $M_i(t)$, (ii) a naive MCMC approach that replaces $X_i(t)$ with $W_i(t)$ and does not correct for measurement error, (iii) the full MCMC approach that jointly corrects for measurement error, and (iv) subgroup-specific functional AFT models fitted using \texttt{mgcv::gam}. To evaluate estimation accuracy across simulation replicates, we computed the Monte Carlo root mean squared error (RMSE) and mean integrated squared error (MISE) for the scalar fixed effects and functional effect respectively. 
More specifically, let $\beta^{(r)}(\mbf{Z}_{2i}, t_l)$ denote the true functional effect for the $i^\text{th}$ subject from the $r^\text{th}$ replicate evaluated at the $l^\text{th}$ time point on a grid where $l \in \{1, \dots, T\}$, and $\widehat{\beta}^{(r)}(\mbf{Z}_{2i}, t_l)$ denote the corresponding posterior mean estimate. We computed the subject-averaged integrated squared error and report the mean integrated squared error (MISE) across replicates. That is, 
\begin{align}
ISE^{(r)} &= \frac{1}{n}\sum^n_{i=1}\frac{1}{T}\sum^{T}_{l=1} (\widehat{\beta}^{(r)}(\mbf{Z}_{2i}, t_l) - \beta^{(r)}(\mbf{Z}_{2i}, t_l))^2\\
MISE &= \frac{1}{R}\sum^{R}_{r=1} ISE^{(r)}. 
\end{align}


\subsection{Results}

\begin{figure}[H]
    \centering
    \includegraphics[width=0.9\linewidth]{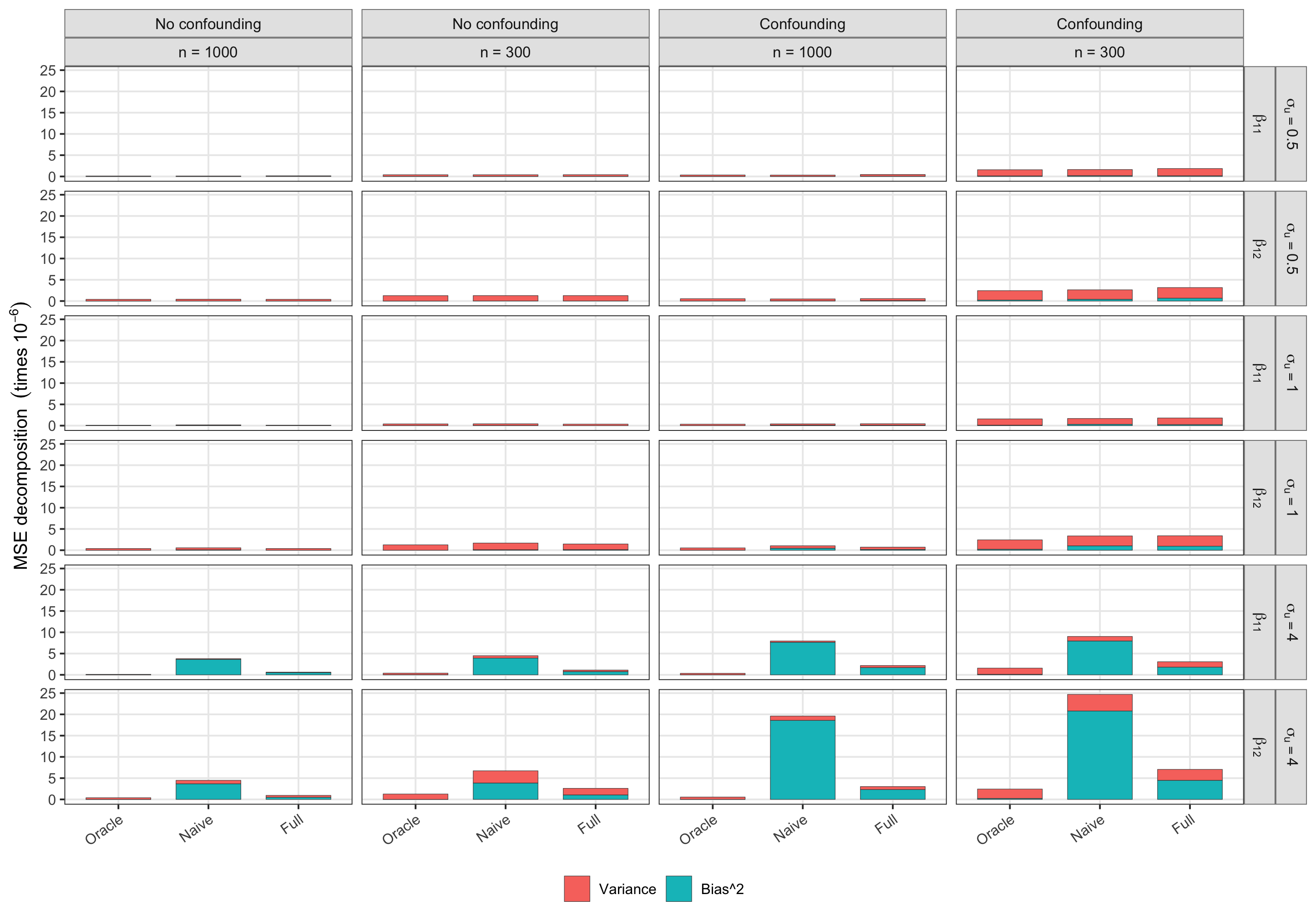}
    \caption{Monte Carlo MSE and squared bias for the fixed effects. Values are reported on the scale of $10^{-6}$. }
    \label{fig:FixEffMSEBias}
\end{figure}

Figure \ref{fig:FixEffMSEBias} summarizes the Monte Carlo mean squared error (MSE) and squared bias for the scalar fixed effects across all simulation scenarios. Under the first simulation setting where the scalar covariates are generated independently of the latent $X_i(t)$, all approaches produce relatively small MSE values, particularly when the measurement error variance is low and the sample size is large. As $\sigma_u$ increases, both the naive and full approaches exhibit increases in MSE and squared bias, although the deterioration is generally less pronounced for the full approach. Overall, the differences between the naive and full approaches remain modest, suggesting that measurement error in the functional covariate has a limited impact on estimation of the scalar regression coefficients when no confounding is present. 

\begin{figure}
    \centering
    \includegraphics[width=0.75\linewidth]{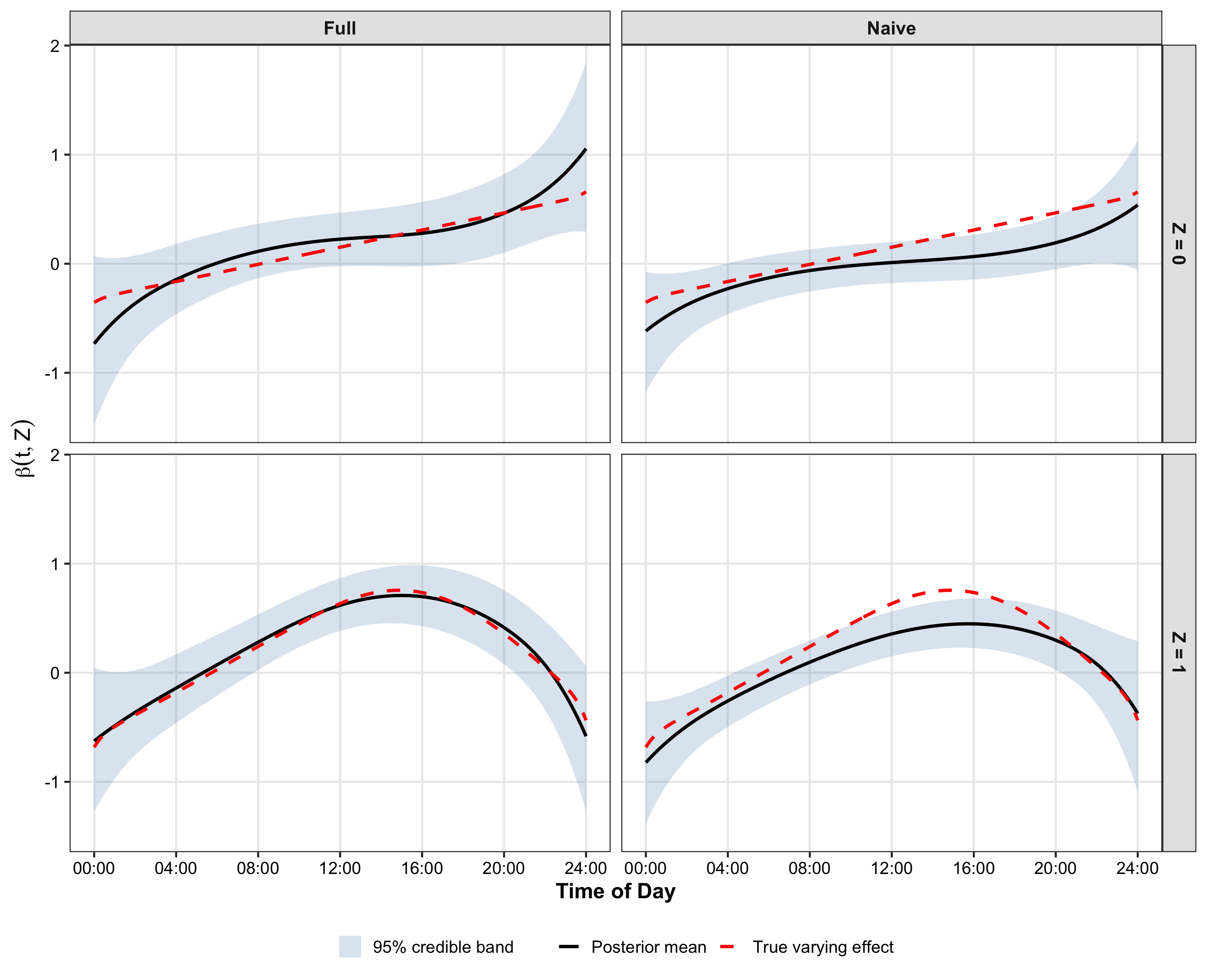}
    \caption{Posterior mean and 95\% credible interval for the estimate of $\beta(Z_{2i}, t)$ based on the full and naive approaches in the main simulation study. Estimates are organized by the binary subgroups defined by $Z_{2i}$. }
    \label{fig:Bestplot}
\end{figure}

We next examine the estimation of the functional varying effect across all competing approaches in Table \ref{tab:beta-mise-combined}. We simulated datasets with $\sigma_\varepsilon = 0.1$ and measurement error variances of $\sigma_u^2 = 0.5^2, 1.0$, and $4.0^2$. At each setting of $\sigma_u$, we compared the MISE performance across different sample sizes, correlation levels in $\rho_u$, and censoring levels. Since the oracle approach assumes the latent functional covariate $X_i(t)$ to be observed, its MISE values remain unchanged across different values of $\sigma_u$ and serve as a benchmark for comparison. Under low measurement error ($\sigma_u = 0.5$), the naive approach achieves MISE values comparable to the oracle approach since the observed proxy $W_i(t)$ is similar to $X_i(t)$. However, as $\sigma_u$ increases, the MISE values from the naive approach increase substantially. 
This behavior is illustrated in Figure \ref{fig:Bestplot}, where the credible intervals corresponding to the naive approach do not fully capture the true effect in some time periods. 
We further observe that increasing the correlation level $\rho_u$ generally leads to larger MISE values for the naive approach, with the deterioration becoming more pronounced as $\sigma_u$ increases. 
In contrast, the full approach that estimates the latent $X_i(t)$ is less sensitive to increasing measurement error and generally outperforms the naive approach under larger values of $\sigma_u$, $\rho_u$, and sample sizes. 
The subgroup-specific general additive models fitted using \texttt{mgcv::gam} consistently exhibits higher MISE values than the Bayesian approaches, which is expected since the subgroup models do not assume a functional varying effect, nor do they explicitly model measurement error in the functional covariate. 
Simulation settings with censoring levels at 80\% showed higher MISE values across all methods, since a greater proportion of survival times were imputed during posterior sampling. Despite this, the full approach continues to outperform the naive method across censoring levels. Together, these results demonstrate that estimation of the functional varying effect without correcting for measurement error can result in biased estimates with larger MISE values, particularly when the measurement error process exhibits high variance and strong temporal correlation. 


We next consider simulations with possible confounding between the scalar and latent functional covariates. Specifically, we induced dependence between $\mbf{Z}_{1i}$ and $X_i(t)$ through the basis coefficients used to construct the latent functional trajectories. Under this setting, the measurement error in the functional covariates may potentially affect estimation of both the scalar fixed effects and the functional varying effect through the induced dependence structure. As shown in Figure \ref{fig:FixEffMSEBias}, the MSE values for the scalar fixed effects are generally higher compared to the previous simulation setting, with the increase drivenprimarily by squared bias rather than variance. The increase in MSE is particularly pronounced for the naive approach under larger values of $\sigma_u$. Decomposition of the MSE shows that the deterioration is primarily from increased bias, while the variance remains comparatively stable. Thus, the induced dependence allows measurement error in the functional covariate to introduce systematic bias into estimation of the scalar covariate fixed effects. 

Table \ref{tab:newMISE_Confounding} summarizes the estimation results of the functional varying effect under the second simulation setting. Compared to the results in Table \ref{tab:beta-mise-combined}, the presence of the scalar covariates $\mbf{Z}_1$ as a confounder led to larger MISE values across all approaches. Investigation of the average squared bias and average variance in the MISE decomposition reveals that the increase is primarily driven by bias, while the variance remains comparatively stable across simulation settings. The effect is particularly pronounced for the naive method, where settings with larger values of $\sigma_u$ show heavily deteriorated bias and MISE values. While the full approach also exhibits larger MISE under increased $\sigma_u$, the deterioration is comparatively milder and indicates how modeling the latent $X_i(t)$ can mitigate the bias introduced by measurement error in $W_i(t)$. As in the first simulation setting, the MISE values for all methods are lower for settings with larger sample sizes. Overall, these results suggest that the presence of measurement error primarily impacts the estimation of functional varying effects through increased bias, with the impact becoming more pronounced in the presence of confounding. 
\begin{table}[H]
\centering
\caption{Monte Carlo MISE for the varying functional effect under different measurement error variances ($\sigma_u$), correlation levels ($\rho_u$), censoring levels, and sample sizes. Results are based on 800 simulated datasets with $\sigma_{\epsilon}=0.1$.}
\label{tab:beta-mise-combined}
\begin{tabular}{llccccc}
\toprule
& & & \multicolumn{3}{c}{40\% Censoring}
& 80\% Censoring \\
\cmidrule(lr){4-6}\cmidrule(lr){7-7}
Method & $n$ & $\sigma_u$
& $\rho_u=0.00$
& $\rho_u=0.25$
& $\rho_u=0.75$
& $\rho_u=0.25$ \\
\midrule

Oracle
& 300
& --
& \multicolumn{3}{c}{0.0248}
& 0.0871 \\
& 1000
& --
& \multicolumn{3}{c}{0.0080}
& 0.0256 \\

\midrule

Naive
& \multirow{3}{*}{300}
& 0.50
& 0.0254
& \textbf{0.0262}
& 0.0318
& \textbf{0.0872} \\
&
& 1.00
& 0.0285
& \textbf{0.0318}
& 0.0619
& \textbf{0.0900} \\
&
& 4.00
& 0.1069
& 0.1423
& 0.2337
& 0.1628 \\

& \multirow{3}{*}{1000}
& 0.50
& 0.0085
& \textbf{0.0091}
& 0.0158
& \textbf{0.0256} \\
&
& 1.00
& 0.0117
& 0.0155
& 0.0499
& \textbf{0.0320} \\
&
& 4.00
& 0.0976
& 0.1344
& 0.2297
& 0.1333 \\

\midrule

Full
& \multirow{3}{*}{300}
& 0.50
& 0.0470
& 0.0463
& 0.0432
& 0.1588 \\
&
& 1.00
& 0.0438
& 0.0424
& \textbf{0.0390}
& 0.1628 \\
&
& 4.00
& \textbf{0.0417}
& \textbf{0.0457}
& \textbf{0.0602}
& \textbf{0.1410} \\

& \multirow{3}{*}{1000}
& 0.50
& 0.0169
& 0.0163
& \textbf{0.0139}
& 0.0397 \\
&
& 1.00
& 0.0148
& \textbf{0.0138}
& \textbf{0.0130}
& 0.0358 \\
&
& 4.00
& \textbf{0.0172}
& \textbf{0.0252}
& \textbf{0.1065}
& \textbf{0.0625} \\

\midrule

\texttt{mgcv::gam}
& 300
& --
& \multicolumn{3}{c}{0.4638}
& 0.4768 \\
& 1000
& --
& \multicolumn{3}{c}{0.4606}
& 0.4594 \\

\bottomrule
\end{tabular}
\end{table}

\begin{table}[H]
\centering
\caption{Simulation results with $\mbf{Z}_1$ as a confounder and varying values of $n$ and $\sigma_u$. }
\label{tab:newMISE_Confounding}
\begin{tabular}{l l c c c c | l l c c c}
\toprule
Method & $n$ & $\sigma_u$ & $\text{ABias}^2$ & $\text{AVar}$ & $\text{MISE}$ & $n$ &$\sigma_u$ &$\text{ABias}^2$ & $\text{AVar}$ & $\text{MISE}$\\
\midrule
Oracle & 300 & -- & 0.0118 & 0.0119 & 0.0237
       & 1000 & -- & 0.0016 & 0.0058 & 0.0074 \\

\midrule

Naive  & 300 & 0.5 & 0.0138 & 0.0114 & \textbf{0.0252}
       & 1000 & 0.5 & 0.0028 & 0.0058 & \textbf{0.0086} \\
       & 300 & 1.0 & 0.0202 & 0.0107 & \textbf{0.0309}
       & 1000 & 1.0 & 0.0095 & 0.0051 & 0.0146 \\
       & 300 & 4.0 & 0.1280 & 0.0066 & 0.1346
       & 1000 & 4.0 & 0.1230 & 0.0027 & 0.1257 \\

\midrule

Full   & 300 & 0.5 & 0.0244 & 0.0122 & 0.0366
       & 1000 & 0.5 & 0.0058 & 0.0086 & 0.0144 \\
       & 300 & 1.0 & 0.0240 & 0.0122 & 0.0362
       & 1000 & 1.0 & 0.0053 & 0.0085 & \textbf{0.0138} \\
       & 300 & 4.0 & 0.0271 & 0.0151 & \textbf{0.0422}
       & 1000 & 4.0 & 0.0131 & 0.0162 & \textbf{0.0293} \\\bottomrule
\end{tabular}
\end{table}

\section{Application}

\begin{figure}[h!]
    \centering
    \includegraphics[width=1\linewidth]{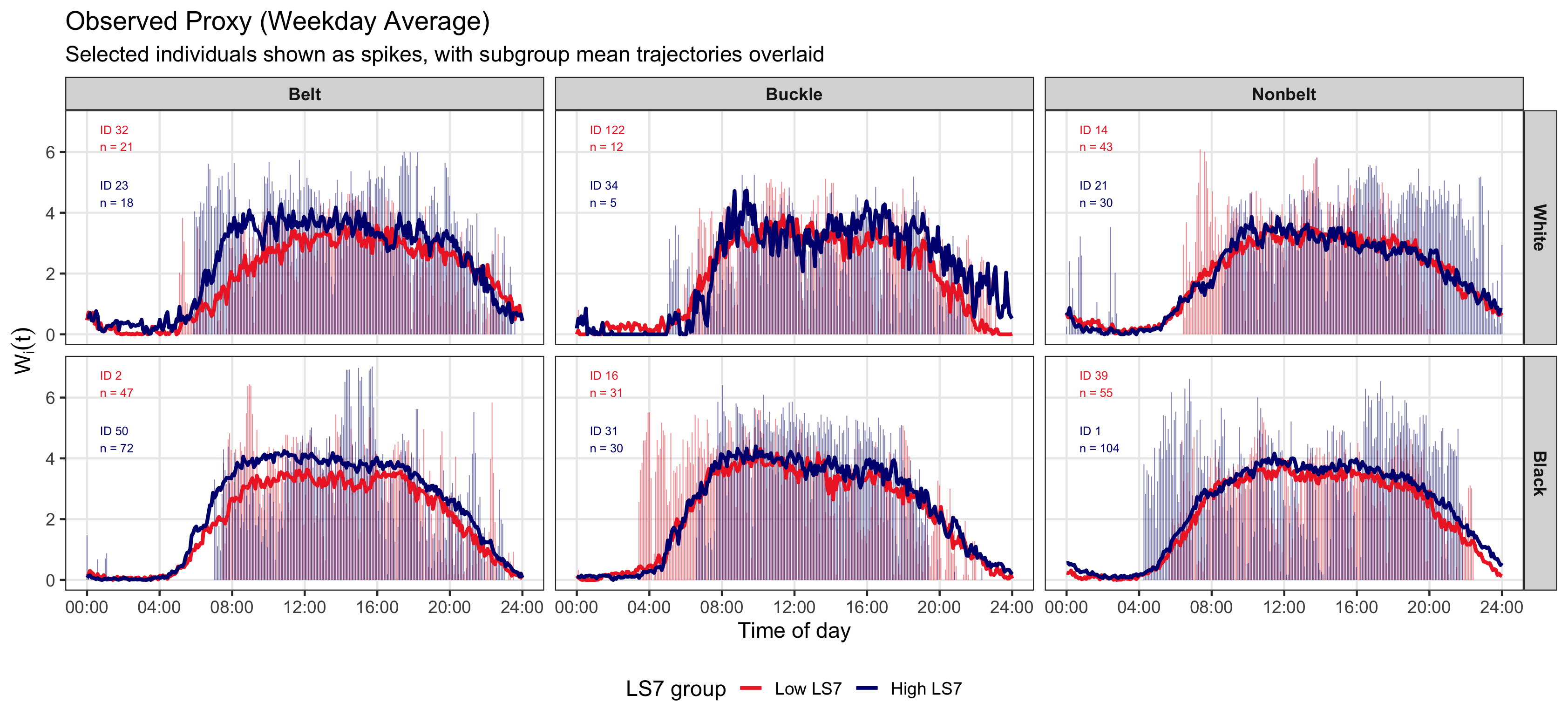}
    \caption{Plot of observed proxy $W_i(t)$ (weekday average log activity count) in the REGARDS cohort. }
    \label{fig:REGARDS_W_ggplot}
\end{figure}

\begin{figure}[h!]
    \centering
    \includegraphics[width=1\linewidth]{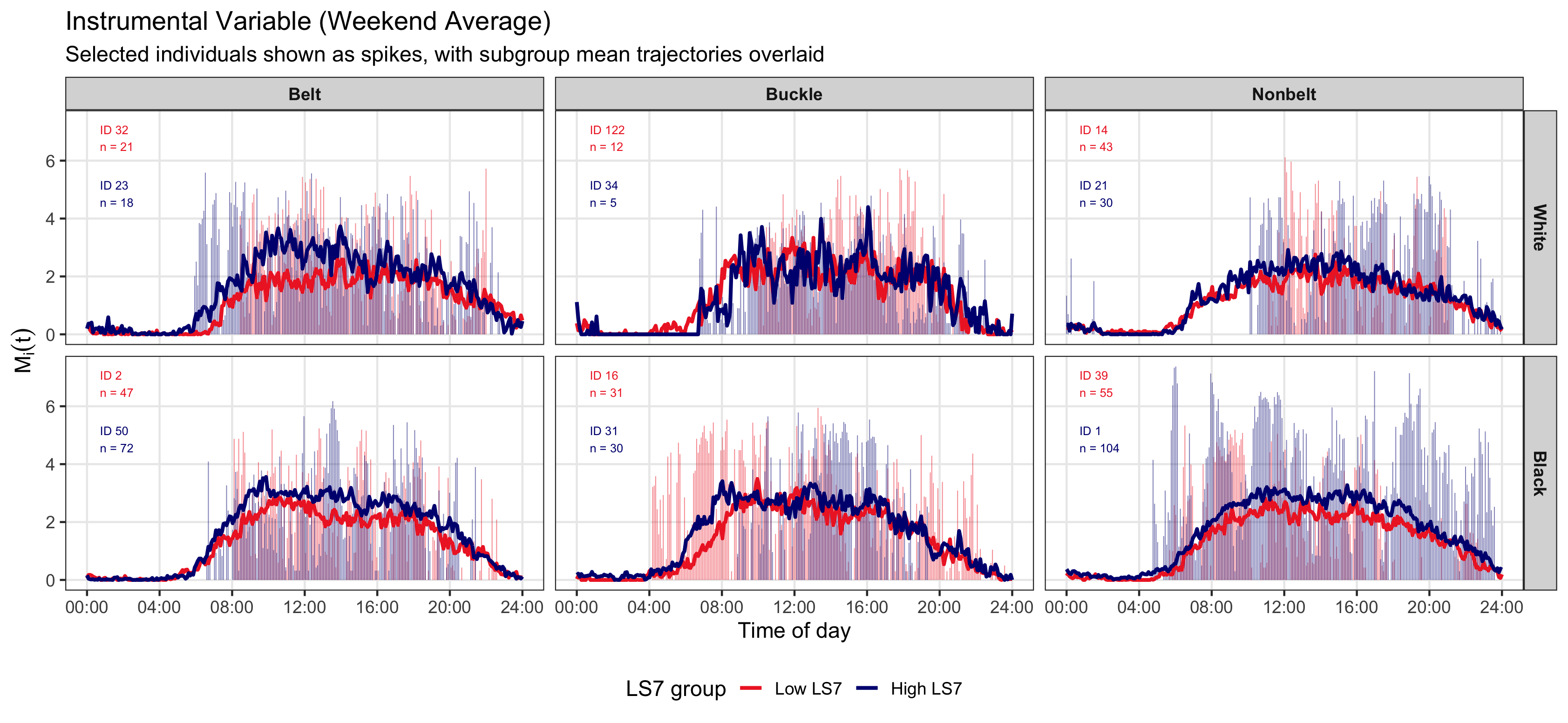}
    \caption{Plot of instrumental variable $M_i(t)$ (weekend average log activity count) in the REGARDS cohort. }
    \label{fig:REGARDS_M_ggplot}
\end{figure}

We applied the proposed Bayesian functional varying-effects AFT model to data from the Reasons for Geographic and Racial Differences in Stroke (REGARDS) study, a national prospective cohort initiated between 2003 and 2007 enrolling 30,239 adults aged 45 and above, with oversampling in the southeastern ``stroke belt'' region to investigate racial and geographic disparities in stroke and cardiovascular disease (\citealt{howard2005reasons}). 
Previous analyses of the REGARDS cohort have primarily used Bayesian hierarchical models to study spatial and demographic heterogeneity in scalar cardiovascular health outcomes (\citealt{tabb2022spatially}). 
Our analyses included 1,268 individuals with complete accelerometer data for a full week. 
Given the distinct roles of race and geographic regions in the REGARDS study, we conducted two separate analyses in which the functional varying effect was allowed to vary with either race or region, along with season and the Life's Simple Seven (LS7) score. 
The LS7 score is a composite measure of behavioral and biological risk factors (e.g. blood pressure, smoking habits, and BMI) related to cardiovascular health, where previous studies of the REGARDS cohort have demonstrated that changes in cardiovascular health are more effectively captured through an aggregate LS7 score compared to its separate behavioral and clinical components (\citealt{garg2018usefulness}, and \citealt{liu2021change}). 
To account for systematic differences in physical activity across days of the week (\citealt{to2022differences}), we treated the log-transformed weekday and weekend average activity binned to five-minute intervals as the observed proxy ($W_i(t)$) and instrumental variable ($M_i(t)$) respectively. 
Figures \ref{fig:REGARDS_W_ggplot} and \ref{fig:REGARDS_M_ggplot} display the weekday and weekend average activity respectively for representative individuals across the subgroup categories of interest, with subgroup mean trajectories overlaid. 
The scalar covariates considered in our analyses include age, gender, cancer history, and the neighborhood socioeconomic score (NSES). 

\newpage

\begin{figure}[h!]
    \centering
    \includegraphics[width=1\linewidth]{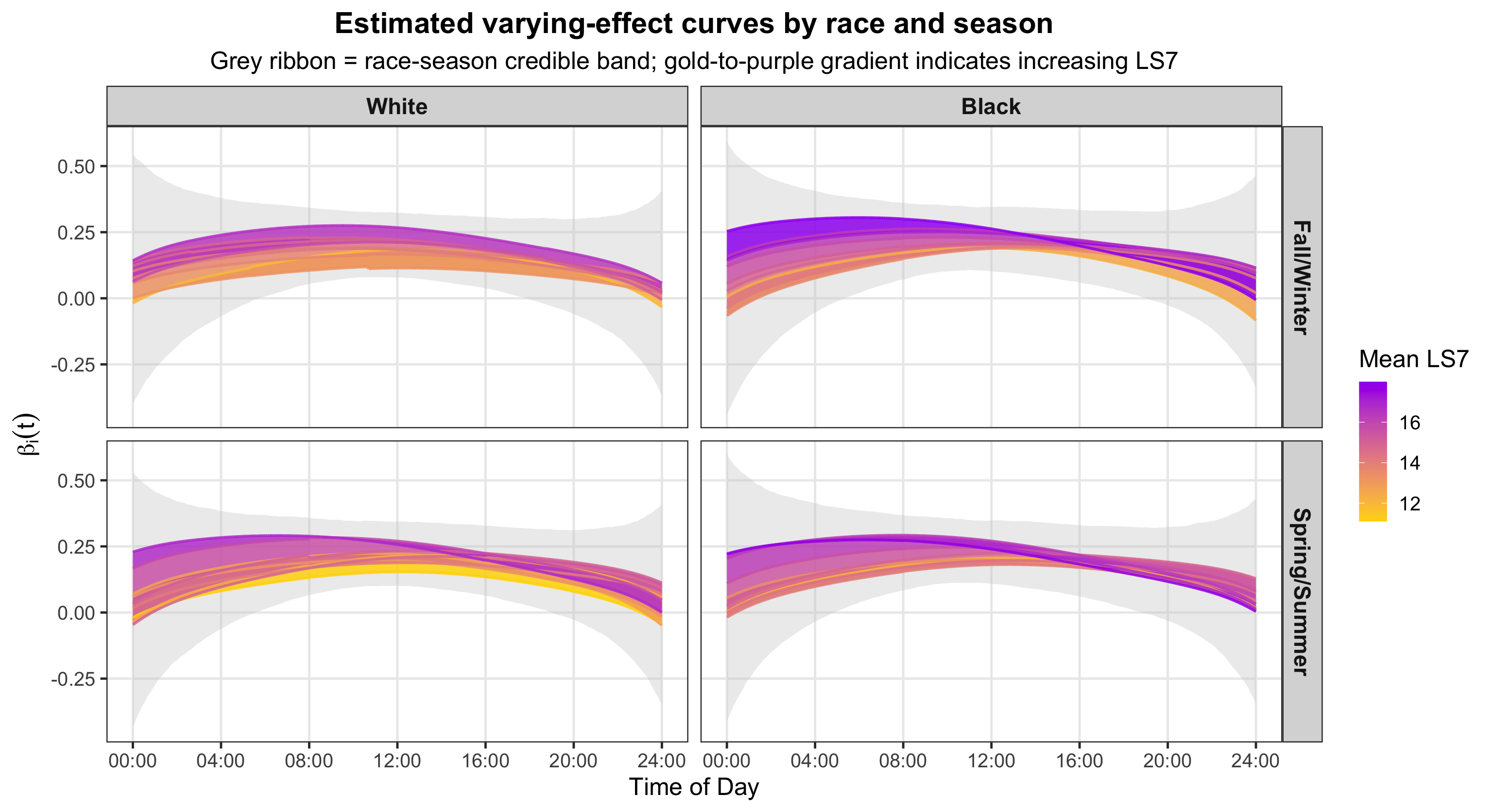}
    \caption{Posterior estimate of functional effect varying by race, season, and LS7. }
    \label{fig:RaceSeasonLS7VarEff}
\end{figure}

We used the priors discussed in Section 2.4 and ran our MCMC for 10,000 iterations discarding 2,000 as burn-in. 
For the race-based analysis, the estimated functional varying effect shows a broadly consistent pattern across the two racial subgroups, with physical activity having the strongest positive association to longer survival times during the morning and mid-day hours (approximately 06:00 to 17:00). 
Figure \ref{fig:RaceSeasonLS7VarEff} illustrates the estimated functional effect $\beta(\mbf{Z}_{2i}, t)$ as a function of race, season, and LS7 score. Across both racial groups and seasonal panels, the estimated curves share similar curvatures and magnitudes. An increasing gradient pattern associated with LS7 score is visible, with higher LS7 scores corresponding to higher values of $\beta(\mbf{Z}_{2i}, t)$ throughout the day. The overlapping credible interval bands across race groups and seasons suggest modest differences between these groupings. Overall, the estimated association between physical activity and survival appears largely consistent across the examined subgroups. 

We next examine the analysis where the functional varying effect assumes heterogeneity based on geographic region, season, and LS7 scores. The posterior estimate of the functional effect shows temporal patterns consistent with those observed in the previous race-based analysis, with effects increasing from early morning and declining later in the day. Across regions, the functional varying effects exhibit similar magnitudes and shapes, with overlapping corresponding credible interval bands. While a modest gradient associated with LS7 score is visible, the overall pattern remains consistent across regions and seasons. These findings suggest that geographic region does not contribute to heterogeneity in the estimated relationship between physical activity and survival time. 
\begin{figure}[h!]
    \centering
    \includegraphics[width=1\linewidth]{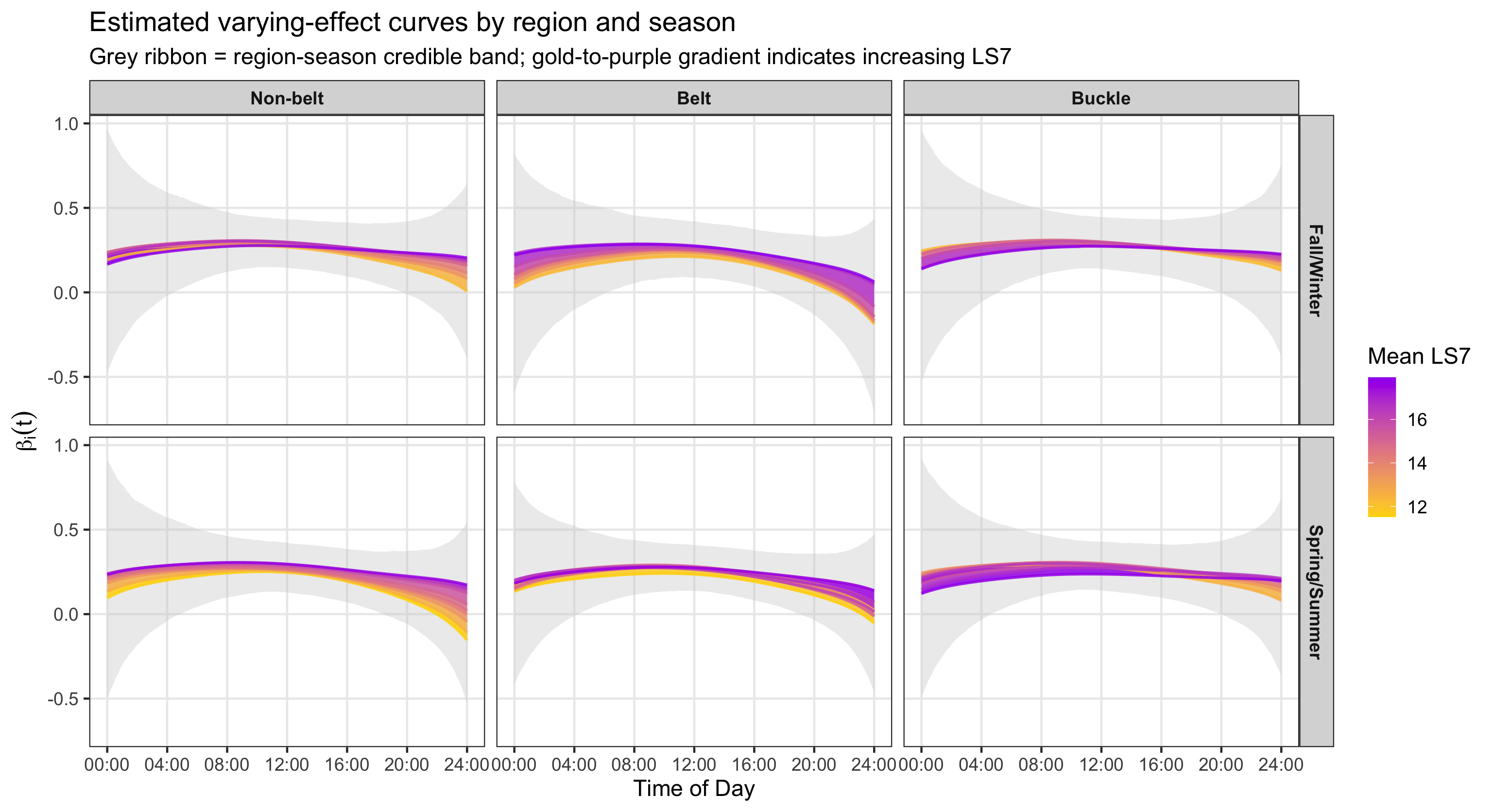}
    \caption{Posterior estimate of functional effect varying by region, seasons, and LS7. }
    \label{fig:RegionSeasonLS7VarEff}
\end{figure}

We evaluated predictive performance using the integrated Brier score (IBS) and concordance C index. Predictions were constructed using both the naive MCMC approach based on the observed proxy $W_i(t)$, and the full MCMC approach using posterior-based recovery of the latent functional covariate $X_i(t)$. As shown in Table \ref{tab:cindex_ibs}, the full approach correcting for measurement error yields the lowest IBS in both the race-and region-based analyses, although improvements relative to the naive approach are modest. A similar ordering is observed for the C-index, with the full MCMC approach achieving the highest concordance, followed by the naive approach and the \texttt{mgcv::gam} model that ignores varying effects. These results indicate that accounting for measurement error improves predictive accuracy, while both Bayesian approaches outperform the subgroup-specific general additive models. 
The scaling function $\delta(t)$ was estimated empirically as $\widehat{\delta}(t) = \frac{\sum^n_{i=1}M_i(t)}{\sum^n_{i=1}W_i(t)}$ and treated as fixed during posterior sampling, consistent with previous work (\citealt{zoh2024bayesian}). 
Table \ref{tab:REGARDS_fixed_effects_combined} summarizes posterior mean estimates and 95\% credible intervals for the error-free scalar covariates in both race- and region-based analyses. Age exhibits a consistent negative association with log-survival time, while the effects of gender, cancer history, and NSES are comparatively modest with credible intervals containing zero. 
\begin{table}[ht]
\centering
\caption{Comparison of concordance index (C-index) and integrated Brier score (IBS) for the race-based and region-based analyses. IBS values are shown in parentheses.}
\label{tab:cindex_ibs}
\begin{tabular}{lcc}
\hline
Method & Race-based & Region-based \\
\hline
Full MCMC & 0.7262 (0.0479) & 0.7396 (0.0494) \\
Naive MCMC using $W_i(t)$         & 0.7195 (0.0491) & 0.7300 (0.0488) \\
\texttt{mgcv::gam} ignoring varying effects                       & 0.6880 (0.0591) & 0.6880 (0.0591) \\
\hline
\end{tabular}
\end{table}



\begin{table}[H]
\centering
\caption{Posterior means and 95\% credible intervals for scalar covariates under race-based and region-based analyses}
\begin{tabular}{lccccc}
\hline
& \multicolumn{2}{c}{Race} & \multicolumn{2}{c}{Region} \\
\cline{2-3} \cline{4-5}
Covariate & Mean & 95\% Credible interval & Mean & 95\% Credible interval \\
\hline
Intercept 
& 2.9235 & (2.4634, 3.3601) 
& 3.0087 & (2.6301, 3.4390) \\

Gender (male) 
& -0.1012 & (-0.2301, 0.0275) 
& -0.1017 & (-0.2301, 0.0228) \\

Age 
& -0.1723 & (-0.2494, -0.1009) 
& -0.1771 & (-0.2489, -0.1089) \\

Cancer (yes) 
& -0.1251 & (-0.3124, 0.0591) 
& -0.1318 & (-0.3156, 0.0509) \\

DASH 
& 0.0351 & (-0.0295, 0.1022) 
& 0.0281 & (-0.0356, 0.0915) \\

NSES 
& 0.0056 & (-0.0072, 0.0187) 
& 0.0052 & (-0.0074, 0.0179) \\

\hline
\end{tabular}
\label{tab:REGARDS_fixed_effects_combined}
\end{table}

\section{Conclusion}

In this work, we proposed a Bayesian scalar-on-function accelerated failure time model that accommodates covariate-dependent functional effects and corrects for measurement error in functional predictors. Our proposed framework integrates Gaussian process single-index structures and instrumental variable-based measurement error modeling, enabling flexible inference on heterogeneous functional associations with survival outcomes, and posterior inference through an efficient Metropolis-within-Gibbs sampler. 

Simulation studies demonstrate that the proposed method achieves higher estimation accuracy than naive approaches that ignore measurement error, particularly under substantial noise in the observed functional covariate. Across a range of scenarios, the model recovers the underlying varying-effect structure, highlighting the importance of jointly modeling measurement error and heterogeneous effects in functional survival analysis. In the REGARDS cohort application, the estimated functional varying effect exhibited broadly similar patterns across race, region, and seasons, with modest variation associated with the Life's Simple 7 score. These findings suggest that the relationship between physical activity and survival is largely consistent across the examined subgroups, while revealing heterogeneity related to cardiovascular health status. 
Compared with subgroup-specific frequentist approaches, the proposed framework provides a unified model that captures subgroup-specific functional effects while maintaining comparable fixed-effect estimates across model specifications. 

Future work may extend the proposed framework in several directions. Alternative representations of the varying functional effect, such as tensor product basis expansions, may proide greater flexibility in capturing interactions between time and scalar covariates. More flexible distributional assumptions, including Dirichlet process mixture-based error models, could improve robustness to departures from Gaussian assumptions. Finally, scalable alternatives such as variational inference or integrated nested Laplace approximations may be advantageous for large-scale applications. 
\section{Acknowledgments}
This research was supported by the National Institutes of Health under grants R01DK136994 and R01DK132385.

\renewcommand{\bibname}{Bibliography}
\phantomsection 
\begin{singlespace}
\bibliographystyle{apacite}
\addcontentsline{toc}{chapter}{\texorpdfstring{\uppercase{Bibliography}}{Bibliography}}
\bibliography{sample.bib}

@article{sarkar2018bayesian,
	author = {Sarkar, Abhra and Pati, Debdeep and Chakraborty, Antik and Mallick, Bani K and Carroll, Raymond J},
	journal = {Journal of the American Statistical Association},
	number = {521},
	pages = {401--416},
	publisher = {Taylor \& Francis},
	title = {Bayesian semiparametric multivariate density deconvolution},
	volume = {113},
	year = {2018}}

@article{neal2011mcmc,
	author = {Neal, Radford M and others},
	journal = {Handbook of markov chain monte carlo},
	number = {11},
	pages = {2},
	title = {MCMC using Hamiltonian dynamics},
	volume = {2},
	year = {2011}}

@book{Carroll2006,
	author = {Carroll, R. J. and Ruppert, D. and Stefanski, L. A. and Crainiceanu, C. M.},
	publisher = {Chapman and Hall},
	title = {Measurement Error in Nonlinear Models: A Modern Perspective, Second Edition},
	year = {2006}}

@article{howard2005reasons,
	author = {Howard, Virginia J and Cushman, Mary and Pulley, LeaVonne and Gomez, Camilo R and Go, Rodney C and Prineas, Ronald J and Graham, Andra and Moy, Claudia S and Howard, George},
	journal = {Neuroepidemiology},
	number = {3},
	pages = {135--143},
	publisher = {Karger Publishers},
	title = {The reasons for geographic and racial differences in stroke study: objectives and design},
	volume = {25},
	year = {2005}}

@article{goldsmith2011penalized,
	author = {Goldsmith, Jeff and Bobb, Jennifer and Crainiceanu, Ciprian M and Caffo, Brian and Reich, Daniel},
	journal = {Journal of Computational and Graphical Statistics},
	number = {4},
	pages = {830--851},
	publisher = {Taylor \& Francis},
	title = {Penalized functional regression},
	volume = {20},
	year = {2011}}

@article{li2010generalized,
	author = {Li, Yehua and Wang, Naisyin and Carroll, Raymond J},
	journal = {Journal of the American Statistical Association},
	number = {490},
	pages = {621--633},
	publisher = {Taylor \& Francis},
	title = {Generalized functional linear models with semiparametric single-index interactions},
	volume = {105},
	year = {2010}}

@article{choi2011gaussian,
	author = {Choi, Taeryon and Shi, Jian Q and Wang, Bo},
	journal = {Journal of Nonparametric Statistics},
	number = {1},
	pages = {21--36},
	publisher = {Taylor \& Francis},
	title = {A Gaussian process regression approach to a single-index model},
	volume = {23},
	year = {2011}}

@article{antoniadis2004bayesian,
	author = {Antoniadis, Anestis and Gr{\'e}goire, G{\'e}rard and McKeague, Ian W},
	journal = {Statistica Sinica},
	pages = {1147--1164},
	publisher = {JSTOR},
	title = {Bayesian estimation in single-index models},
	year = {2004}}

@article{zoh2024bayesian,
	author = {Zoh, Roger S and Luan, Yuanyuan and Xue, Lan and Allison, David B and Tekwe, Carmen D},
	journal = {Statistics in Medicine},
	number = {21},
	pages = {4043--4054},
	publisher = {Wiley Online Library},
	title = {A Bayesian semi-parametric scalar-on-function regression with measurement error using instrumental variables},
	volume = {43},
	year = {2024}}

@article{maity2020integration,
	author = {Arnab Kumar Maity and Raymond J. Carroll and Bani K. Mallick},
	issn = {00359254, 14679876},
	journal = {Journal of the Royal Statistical Society. Series C (Applied Statistics)},
	number = {5},
	pages = {pp. 1577--1595},
	publisher = {[Royal Statistical Society, Oxford University Press]},
	title = {Integration of survival and binary data for variable selection and prediction: a Bayesian approach},
	urldate = {2025-03-06},
	volume = {68},
	year = {2019}}

@article{muller2005generalized,
	author = {M{\"u}ller, Hans-Georg and Stadtm{\"u}ller, Ulrich},
	title = {Generalized functional linear models},
	year = {2005}}

@article{mclean2014functional,
	author = {McLean, Mathew W and Hooker, Giles and Staicu, Ana-Maria and Scheipl, Fabian and Ruppert, David},
	journal = {Journal of Computational and Graphical Statistics},
	number = {1},
	pages = {249--269},
	publisher = {Taylor \& Francis},
	title = {Functional generalized additive models},
	volume = {23},
	year = {2014}}

@article{bhattacharya2016fast,
	author = {Bhattacharya, Anirban and Chakraborty, Antik and Mallick, Bani K},
	journal = {Biometrika},
	pages = {asw042},
	publisher = {Oxford University Press},
	title = {Fast sampling with Gaussian scale mixture priors in high-dimensional regression},
	year = {2016}}

@article{luan2023scalable,
	author = {Luan, Yuanyuan and Zoh, Roger S and Cui, Erjia and Lan, Xue and Jadhav, Sneha and Tekwe, Carmen D},
	journal = {arXiv preprint arXiv:2305.12624},
	title = {Scalable regression calibration approaches to correcting measurement error in multi-level generalized functional linear regression models with heteroscedastic measurement errors},
	year = {2023}}

@article{tanner1987calculation,
	author = {Tanner, Martin A and Wong, Wing Hung},
	journal = {Journal of the American statistical Association},
	number = {398},
	pages = {528--540},
	publisher = {Taylor \& Francis},
	title = {The calculation of posterior distributions by data augmentation},
	volume = {82},
	year = {1987}}

@article{tabb2022spatially,
	author = {Tabb, Loni Philip and Roux, Ana V Diez and Barber, Sharrelle and Judd, Suzanne and Lovasi, Gina and Lawson, Andrew and McClure, Leslie A},
	journal = {Spatial and spatio-temporal epidemiology},
	pages = {100473},
	publisher = {Elsevier},
	title = {Spatially varying racial inequities in cardiovascular health and the contribution of individual-and neighborhood-level characteristics across the United States: The REasons for geographic and racial differences in stroke (REGARDS) study},
	volume = {40},
	year = {2022}}

@article{wan2017simulating,
	author = {Wan, Fei},
	journal = {Statistics in medicine},
	number = {5},
	pages = {838--854},
	publisher = {Wiley Online Library},
	title = {Simulating survival data with predefined censoring rates for proportional hazards models},
	volume = {36},
	year = {2017}}

@article{parker2023bayesian,
	author = {Parker, Paul A and Holan, Scott H},
	journal = {Biometrics},
	number = {2},
	pages = {1397--1408},
	publisher = {Oxford University Press},
	title = {A Bayesian functional data model for surveys collected under informative sampling with application to mortality estimation using NHANES},
	volume = {79},
	year = {2023}}

@article{ghosal2023functional,
	author = {Ghosal, Rahul and Matabuena, Marcos and Zhang, Jiajia},
	journal = {arXiv preprint arXiv:2302.07340},
	title = {Functional proportional hazards mixture cure model and its application to modelling the association between cancer mortality and physical activity in NHANES 2003-2006},
	year = {2023}}

@article{shi2005hierarchical,
	author = {Shi, Jian Qing and Murray-Smith, Roderick and Titterington, D Michael},
	journal = {Statistics and computing},
	number = {1},
	pages = {31--41},
	publisher = {Springer},
	title = {Hierarchical Gaussian process mixtures for regression},
	volume = {15},
	year = {2005}}

@article{sinha2010semiparametric,
	author = {Sinha, Samiran and Mallick, Bani K and Kipnis, Victor and Carroll, Raymond J},
	journal = {Biometrics},
	number = {2},
	pages = {444--454},
	publisher = {Oxford University Press},
	title = {Semiparametric Bayesian analysis of nutritional epidemiology data in the presence of measurement error},
	volume = {66},
	year = {2010}}

@article{gencer2025lung,
	author = {Gencer, G{\"u}lcan},
	journal = {T{\"u}rkiye Klinikleri. Tip Bilimleri Dergisi},
	number = {1},
	pages = {8--16},
	publisher = {Ortadogu Reklam Tanitim Yayincilik Turizm Egitim Insaat Sanayi ve Ticaret AS},
	title = {Lung Cancer Survival Analysis: A Comparative Evaluation of Cox Proportional Hazards and Accelerated Failure Time Models: An Analytical Study},
	volume = {45},
	year = {2025}}

@article{qian2025functional,
	author = {Qian, Weijia and Cui, Erjia and Brooks-Russell, Ashley and Wrobel, Julia},
	journal = {arXiv preprint arXiv:2510.22343},
	title = {Functional Accelerated Failure Time Models for Predicting Time Since Cannabis Use},
	year = {2025}}

@article{liu2025efficient,
	author = {Liu, Changyu and Su, Wen and Liu, Kin-Yat and Yin, Guosheng and Zhao, Xingqiu},
	journal = {Biometrika},
	number = {2},
	pages = {asaf011},
	publisher = {Oxford University Press},
	title = {Efficient estimation for functional accelerated failure time models},
	volume = {112},
	year = {2025}}

@article{wang2021semiparametric,
	author = {Wang, Ching-Yun and Song, Xiao},
	journal = {Biometrics},
	number = {2},
	pages = {561--572},
	publisher = {Oxford University Press},
	title = {Semiparametric regression calibration for general hazard models in survival analysis with covariate measurement error; surprising performance under linear hazard},
	volume = {77},
	year = {2021}}

@article{deshpande2020vcbart,
	author = {Deshpande, Sameer K and Bai, Ray and Balocchi, Cecilia and Starling, Jennifer E and Weiss, Jordan},
	journal = {arXiv preprint arXiv:2003.06416},
	pages = {32--33},
	publisher = {Technical report},
	title = {VCBART: Bayesian trees for varying coefficients},
	volume = {2},
	year = {2020}}

@article{guhaniyogi2022distributed,
	author = {Guhaniyogi, Rajarshi and Li, Cheng and Savitsky, Terrance D and Srivastava, Sanvesh},
	journal = {Journal of machine learning research},
	number = {84},
	pages = {1--59},
	title = {Distributed Bayesian varying coefficient modeling using a Gaussian process prior},
	volume = {23},
	year = {2022}}

@article{hastie1993varying,
	author = {Hastie, Trevor and Tibshirani, Robert},
	journal = {Journal of the Royal Statistical Society Series B: Statistical Methodology},
	number = {4},
	pages = {757--779},
	publisher = {Oxford University Press},
	title = {Varying-coefficient models},
	volume = {55},
	year = {1993}}

@article{wu2010varying,
	author = {Wu, Yichao and Fan, Jianqing and M{\"u}ller, Hans-Georg},
	title = {Varying-coefficient functional linear regression},
	year = {2010}}

@article{liu2021change,
	author = {Liu, Chelsea and Roth, David L and Gottesman, Rebecca F and Sheehan, Orla C and Blinka, Marcela D and Howard, Virginia J and Judd, Suzanne E and Cushman, Mary},
	journal = {Stroke},
	number = {3},
	pages = {878--886},
	publisher = {Lippincott Williams \& Wilkins Hagerstown, MD},
	title = {Change in life's simple 7 measure of cardiovascular health after incident stroke: the REGARDS study},
	volume = {52},
	year = {2021}}

@article{garg2018usefulness,
	author = {Garg, Parveen K and O'Neal, Wesley T and Ogunsua, Adedotun and Thacker, Evan L and Howard, George and Soliman, Elsayed Z and Cushman, Mary},
	journal = {The American journal of cardiology},
	number = {2},
	pages = {199--204},
	publisher = {Elsevier},
	title = {Usefulness of the American Heart Association's Life Simple 7 to predict the risk of atrial fibrillation (from the REasons for Geographic And Racial Differences in Stroke [REGARDS] Study)},
	volume = {121},
	year = {2018}}

@article{spiegelman1997regression,
	author = {Spiegelman, Donna and McDermott, Aidan and Rosner, Bernard},
	journal = {The American journal of clinical nutrition},
	number = {4},
	pages = {1179S--1186S},
	publisher = {Elsevier},
	title = {Regression calibration method for correcting measurement-error bias in nutritional epidemiology},
	volume = {65},
	year = {1997}}

@article{to2022differences,
	author = {To, Quyen G and Stanton, Robert and Schoeppe, Stephanie and Doering, Thomas and Vandelanotte, Corneel},
	journal = {Preventive Medicine Reports},
	pages = {101892},
	publisher = {Elsevier},
	title = {Differences in physical activity between weekdays and weekend days among US children and adults: Cross-sectional analysis of NHANES 2011--2014 data},
	volume = {28},
	year = {2022}}
\end{singlespace}
\end{document}